\documentclass[apj]{emulateapj}

\usepackage{dcolumn}
\usepackage{amsmath,amssymb,amsthm,paralist}
\usepackage{graphicx}
\usepackage{color}
\usepackage{appendix}
\usepackage{subfigure}
\usepackage{longtable}
\usepackage{textcomp}
\usepackage{url}
\usepackage{verbatim}
\usepackage{lineno}
\usepackage{bm}

\begin{document}


\title{Consistent Modeling of GS 1826-24 X-Ray Bursts for Multiple
Accretion Rates Demonstrates
the Possibility to Constrain $rp$-Process Reaction Rates}



\author{Zach~Meisel}
\email[]{meisel@ohio.edu}
\affiliation{Institute of Nuclear \& Particle Physics, Department of Physics \& Astronomy, Ohio University, Athens, Ohio 45701, USA}


\begin{abstract}
 Type-I X-ray burst light curves encode unique information about the
 structure of accreting neutron stars and the nuclear reaction rates
 of the $rp$-process that powers bursts. Using the first model
 calculations of hydrogen/helium burning bursts for a large range
 of astrophysical conditions performed with the code {\tt MESA},
 this work shows that simultaneous model-observation comparisons for
 bursts from several accretion rates $\dot{M}$ are required to
 remove degeneracies in astrophysical conditions that otherwise
 reproduce bursts for a single $\dot{M}$ and that such consistent
 multi-epoch modeling could possibly limit the
 $^{15}\rm{O}(\alpha,\gamma)^{19}\rm{Ne}$ reaction rate. Comparisons
 to the year 1998, 2000, and 2007 bursting epochs of the neutron
 star GS 1826-24 show that $\dot{M}$ must be larger than previously
 inferred and that the shallow heating in this source must be below
 0.5~MeV/u, providing a new method to constrain the shallow heating
 mechanism in the outer layers of accreting neutron stars. 
 Features of the light curve rise are used to demonstrate
 that a lower-limit could likely be placed on the 
 $^{15}\rm{O}(\alpha,\gamma)$ reaction rate,
 demonstrating the possibility of constraining nuclear reaction rates
 with X-ray burst light curves.
\end{abstract}


\maketitle

\section{Introduction}

Type-I X-ray bursts, periodic thermonuclear explosions driven and
fueled by hydrogen and/or helium rich mixtures siphoned from a
binary companion, provide unique insight into the structure and
dense matter of the neutron stars that host
them~\citep{Lamb78,Joss78,Scha98,Pari13,Arco17}. X-ray burst models require
detailed input regarding the compositional and thermal structure
of the neutron star envelope, as well as well over a thousand
reaction rates involving more than three hundred
nuclides~\citep{Wall81,Woos04,Scha01,Fisk08,Jose10}. Many important calculation 
inputs,
such as the accretion rate $\dot{M}$ and nuclear reactions rates of the 
rapid proton-capture ($rp$)-process powering bursts,
have distinctive influences on the calculation
results~\citep{Woos04,Pari08,Pari09,Lamp16,Cybu16,Scha17}. 
This enables model-observation
comparisons to determine unique solutions, resulting
in astrophysical constraints on an X-ray bursting
object~\citep{Hege07,Gall17,John18}.

The consistency of the X-ray burster GS 1826-24~\citep{Gall04,Gall08}
and its ``textbook" behavior~\citep{Bild00} have made it the primary
target of past model-observation
comparisons~\citep{Hege07,Gall17,Zamf12}. To date, all of this
pioneering work has been performed using the multizone astrophysical
modeling code {\tt KEPLER}~\citep{Weav78,Woos04}, aside from initial
proof-of-principle calculations performed with the 
open-source multizone stellar evolution code {\tt
MESA}~\citep{Paxt15} and a simple ignition model use to predict burst
recurrence time~\citep{Gall04}. These {\tt KEPLER} model-observation
comparisons
constrained the astrophysical conditions for GS 1826-24 by
reproducing the recurrence time between bursts $\Delta t_{\rm{rec}}$ for several
$\dot{M}$ and the average burst light
curve for a single $\dot{M}$. 
However, simultaneous light curve comparisons for
a consistently modeled range of $\dot{M}$ that approximates the
observed $\dot{M}$ variation have not yet been performed. The peril
in this approach is that the light curve shape from models is known
to vary over the range of $\dot{M}$ similar to that inferred from
observations of
hydrogen/helium-burning Type-I X-ray bursts~\citep{Lamp16}.

Furthermore, the sensitivity of models to varied nuclear reaction
rates has not yet been accounted for in model-observation
comparisons, though some rates are known to substantially impact
model calculations. In particular, the reaction rate
$^{15}\rm{O}(\alpha,\gamma)^{19}\rm{Ne}$ has been shown to alter
$\Delta t_{\rm{rec}}$ beyond observational uncertainties and to
modify the light curve shape much more than the natural variations
observed for GS 1826-24 over its regular bursting
epochs~\citep{Fisk07,Cybu16}.

Here, the first consistent comparison to X-ray burst light curves
for a bursting source over a range of $\dot{M}$ is used to
demonstrate that multi-epoch reproduction is required to remove
degeneracies in astrophysics model parameters  and achieve
tighter astrophysical constraints than previously possible. These
calculations, the first to model hydrogen/helium-burning bursts for
a large range of input conditions with {\tt MESA}, demonstrate that
model-observation comparisons can place tight constraints on the
strength of shallow heating $Q_{\rm{b}}$ in the accreted neutron star
outer layers and that GS 1826-24 has a higher $\dot{M}$ than
previously inferred from models. Additionally, the possibility to
constrain
nuclear reaction rates in the $rp$-process with X-ray burst light
curves is demonstrated by showing that a lower-limit could likely be placed
on the $^{15}\rm{O}(\alpha,\gamma)$
reaction rate. A follow-up paper will compare results from the full grid
of model calculations used for this work 
to results from a similar grid of calculations
performed with {\tt KEPLER}~\citep{Lamp16}. A previous paper featured
{\tt MESA} X-ray burst ash abundances~\citep{Meis17}.

\section{Model Calculations}

Type-I X-ray burst model calculations were performed with the
one-dimensional stellar evolution code {\tt
MESA} version 9793. The numerical approach and
physics models adopted in {\tt MESA} are detailed in the associated
instrumentation papers~\citep{Paxt11,Paxt13,Paxt15,Paxt18}. Here the most
pertinent details for this work are summarized. The neutron star envelope
is 0.01~km-thick, with an inner boundary of neutron star mass 
$M_{\rm{NS}}=1.4~M_{\odot}$ and radius
$R_{\rm{NS}}=11.2$~km, comprised initially of 70\% hydrogen, 28\% helium, and 2\%
metals, by mass, using the solar metallicity $Z$ of
\citet{Grev98}. The envelope is discretized into
$\sim1000$~zones, which adapts during the calculation, where the
local gravity in a zone is corrected for general relativity effects
using a post-Newtonian correction. Convection is approximated using
the mixing length theory of \citet{Heny65} and is
time-dependent~\citep{Paxt11}. 
Accretion is
achieved by adding a small amount of mass to the model's outer
layers and re-adjusting the stellar structure~\citep{Paxt11}. The
spatial and time resolution are adaptive, where the {\tt MESA}
settings {\tt varcontrol\_target=1d-3} and {\tt
mesh\_delta\_coeff=1.0}~\citep{Paxt13} were chosen after tests for
convergence of the light curve shape and $\Delta t_{\rm{rec}}$
in
which {\tt varcontrol\_target} from $10^{-4}-10^{-2}$
{\tt mesh\_delta\_coeff} from $0.5-2.0$ were investigated. 
Here convergence means that the mean light curve and recurrence time 
changed $<<1\sigma$ for finer spatial and/or time resolution
settings. 
The
nuclear reaction network includes the 304 isotopes of
\citet{Fisk08} using reaction rates from the REACLIB V2.2
library~\citep{Cybu10}.

\begin{figure}[t]
\begin{center}
\includegraphics[width=1.0\columnwidth,angle=0]{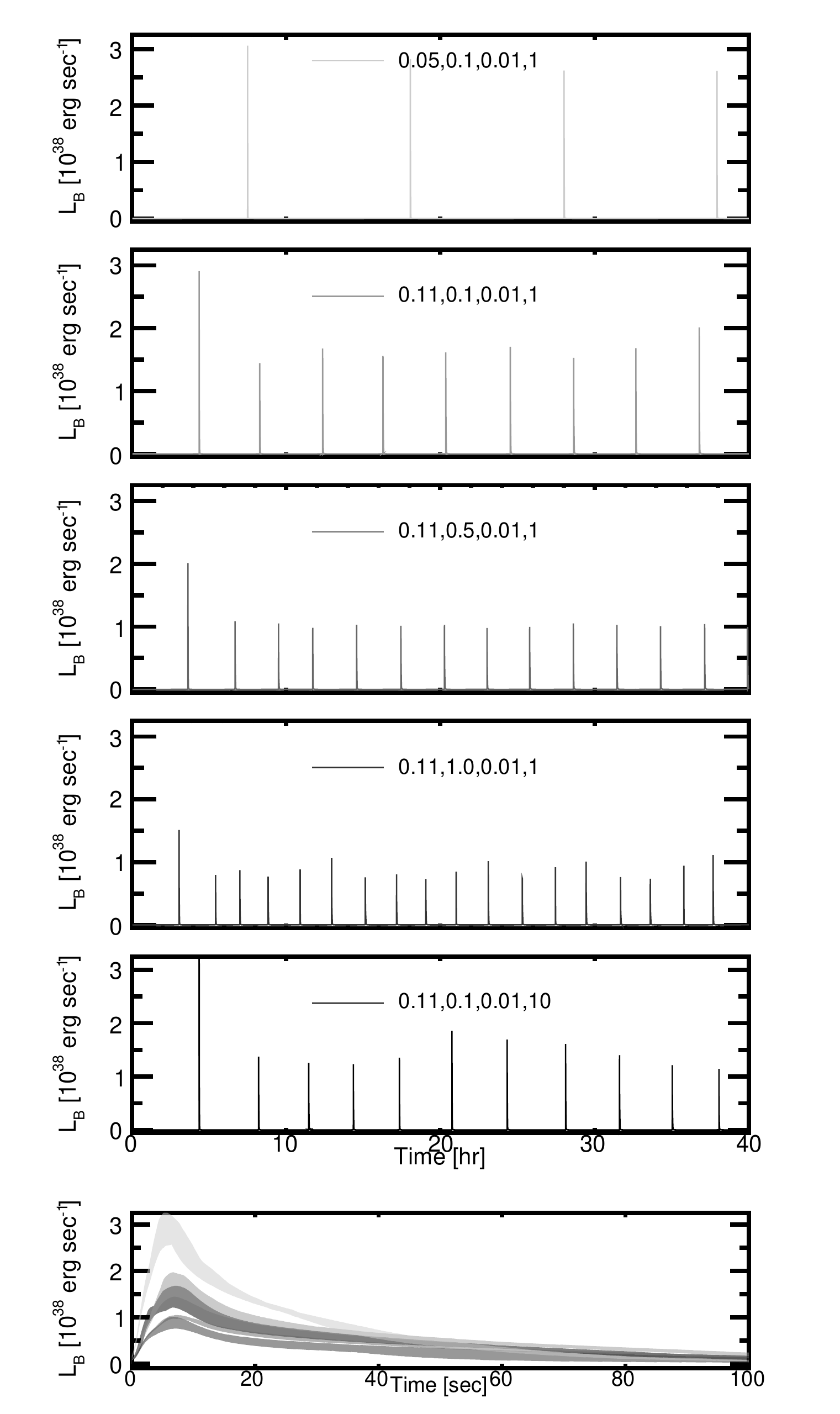}
\caption{Bolometric luminosity over time (not
redshifted) for example {\tt MESA}
calculations. The legends indicate, in order, $\dot{M}$, $Q_{\rm{b}}$, $Z$, and
$R$. $X=0.7$ for each of these models. The
bottom panel shows the average light curve, excluding the first
burst, for all bursts in a sequence for a set of conditions, where
the color of the band matches the color of the corresponding burst
sequence above.
\label{LCexamples}}
\end{center}
\end{figure}

Sequences of X-ray bursts were simulated for 84 different sets of
initial conditions, where models differed in $\dot{M}$,
$Q_{\rm{b}}$, $Z$, hydrogen mass fraction $X$, 
and a reduction factor for the
$^{15}\rm{O}(\alpha,\gamma)$ reaction rate $R$. The number of bursts $N$
belonging to a sequence varies, mostly due to numerical and practical
challenges. Example burst sequences are shown in
Fig.~\ref{LCexamples}. 	
Simulations were performed in sets of three
for $\dot{M}$, with a low, medium, and high multiple of the
Eddington accretion rate $\dot{M}_{\rm{E}}=1.75\times10^{-8}
M_{\odot}$/yr~\citep{Scha99}. 
A low set,
$\dot{M}=0.05, 0.07, 0.08 \dot{M}_{\rm{E}}$, matched observed $\dot{M}$ for
the year 1998, 2000, and 2007 bursting epochs of
GS 1826-24~\citep{Gall08}. A high set, $\dot{M}=0.11, 0.15,
0.17 \dot{M}_{\rm{E}}$,
employed $\dot{M}$ used in the proof-of-principle X-ray burst
calculations of \citet{Paxt15} for the highest $\dot{M}$ 
and then reduced this by the ratio of $\dot{M}$ for the observation
epochs. For example,
$0.17~\dot{M}_{\rm{E}}\times(0.05)/(0.08)\approx0.11~\dot{M}_{\rm{E}}$. 
$Q_{\rm{b}}=0.1, 0.5, 1.0$~MeV/u were used to mimic the
shallow heating of unknown origin that is thought to operate in the
outer layers of accreting neutron stars~\citep{Brow09,Keek17}, where
the lower limit is on the order expected from accretion-induced
reactions in the accreted neutron star crust~\citep{Gupt07,Meis16} and 
the upper limit is on the order of typical shallow heating
inferred from observations of neutron star cooling
after accretion turnoff~\citep{Brow09,Turl15}. In {\tt MESA} this is
achieved by fixing the luminosity of the base of the envelope, so
that the base luminosity depended on $Q_{\rm{b}}$ and $\dot{M}$ of
the model. $Z=0.01, 0.02$ were used to investigate the solar $Z$
favored by previous investigations of GS 1826-24~\citep{Gall04,Hege07}
and a slight reduction from that value. 
$X=0.7$ and helium mass fraction $Y=0.28$ were used for most
simulations, while $X=0.75$ and $Y=0.23$ were employed for a set of
simulations with $Q_{\rm{b}}=0.1$ and $1.0$~MeV/u for each
$\dot{M}$.
$R=1, 5, 10$ were used as $R=10$ is roughly the experimental 
lower-limit for the
$^{15}\rm{O}(\alpha,\gamma)$ rate uncertainty~\citep{Tan07,Davi11}
and rate increases have not been found to impact the burst light
curve~\citep{Cybu16}. Strictly speaking, using a
rate scaling factor is a simplification as compared to using upper
and lower rate limits based on experimental uncertainties. However,
the present rate uncertainty is generally a factor of 10 or larger
in the temperature range of interest~\citep{Davi11}, namely from the onset of
hot CNO cycle burning to break-out, $\sim0.1-0.5$~GK~\citep{Wies99}.
Furthermore, a rate scaling factor enables a more direct comparison
to similar calculations performed in the past with the codes {\tt
KEPLER} and {\tt AGILE}~\citep{Cybu16,Fisk07}.

\vspace{1cm}
\section{Light Curve Construction}

\begin{figure}[htp]
 \begin{center}
  \subfigure[Original.]{\label{fig:Original}\includegraphics[width=1.0\columnwidth]{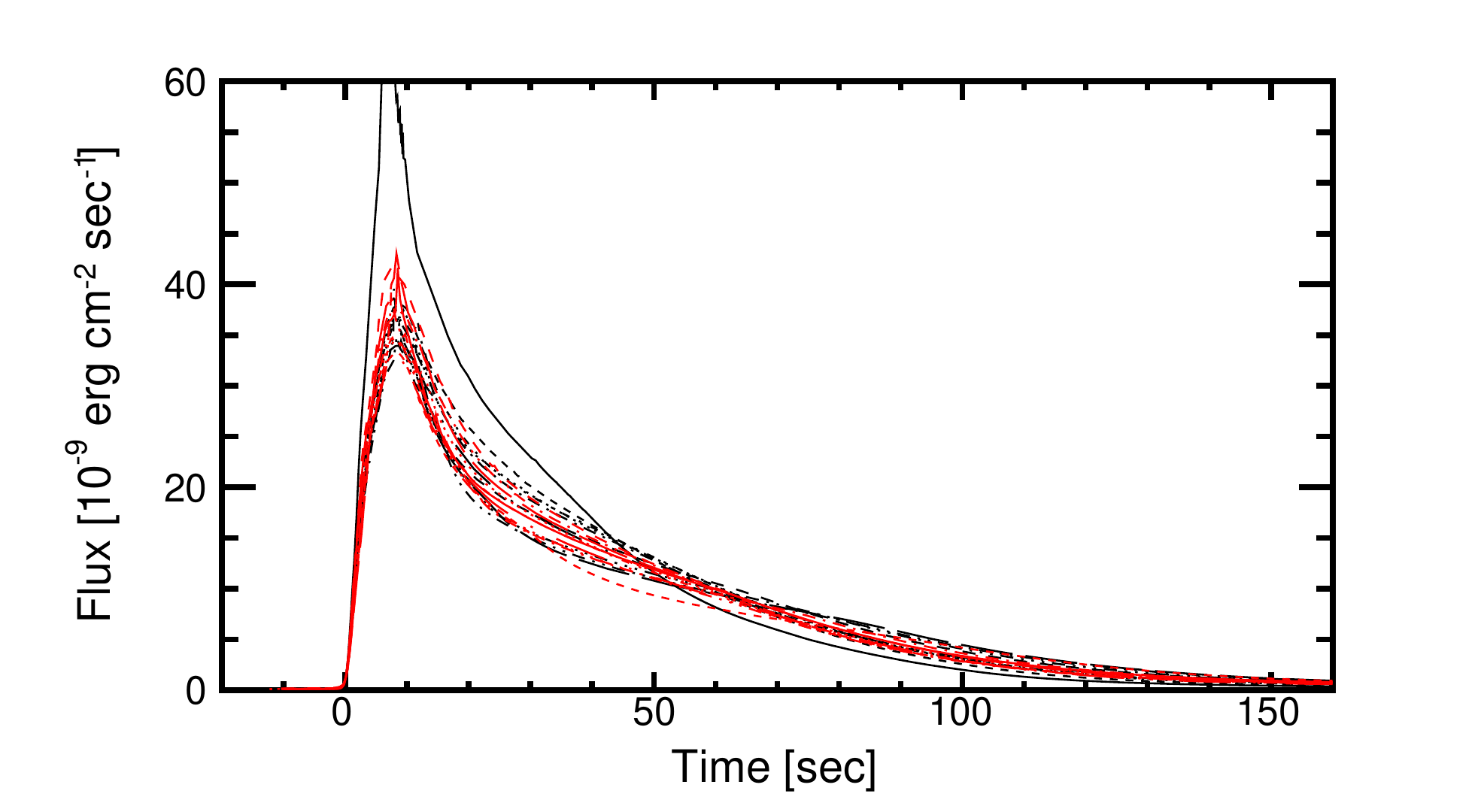}}
  \subfigure[Smoothed.]{\label{fig:Smoothed}\includegraphics[width=1.0\columnwidth]{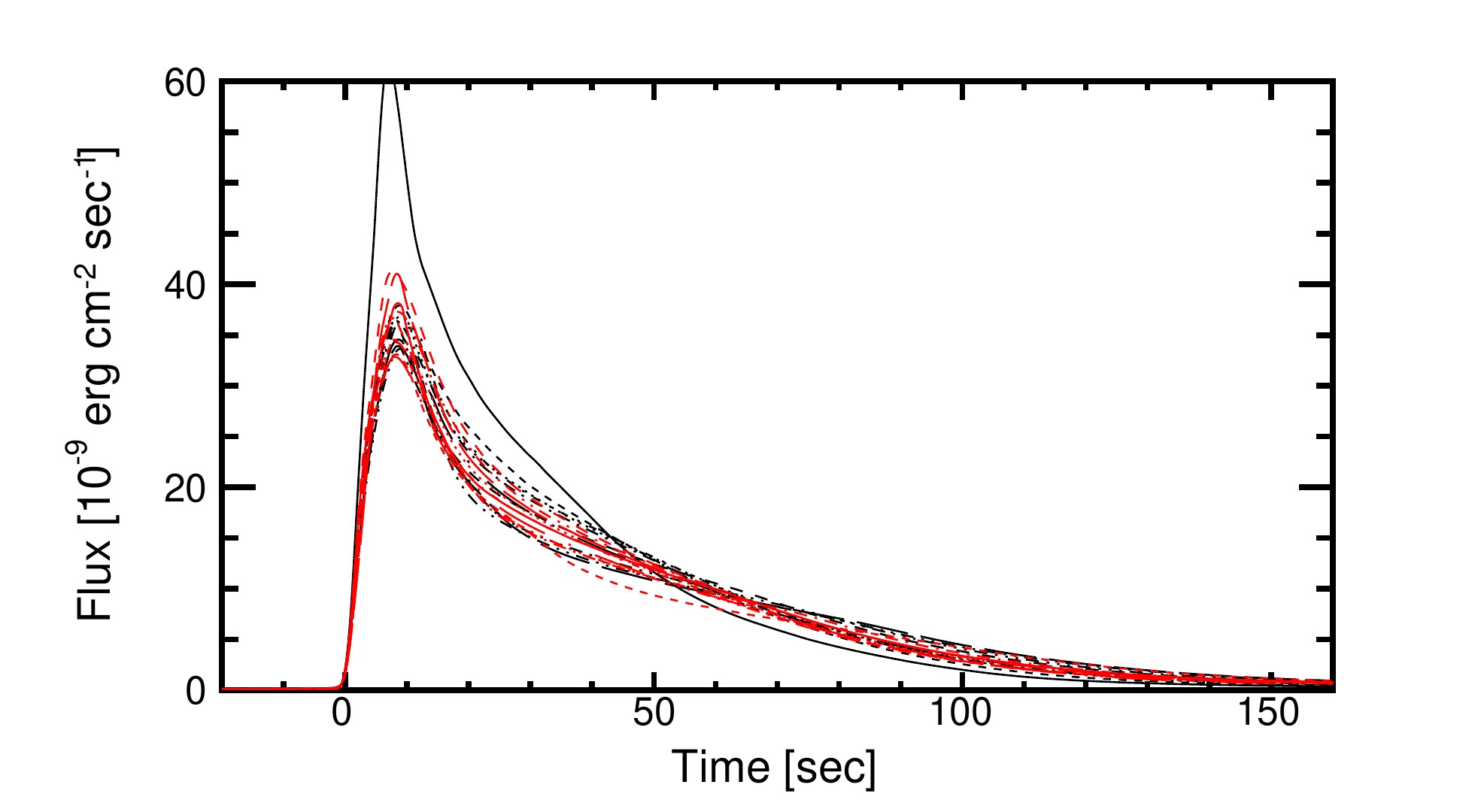}}
  \subfigure[Averaged.]{\label{fig:Averaged}\includegraphics[width=1.0\columnwidth]{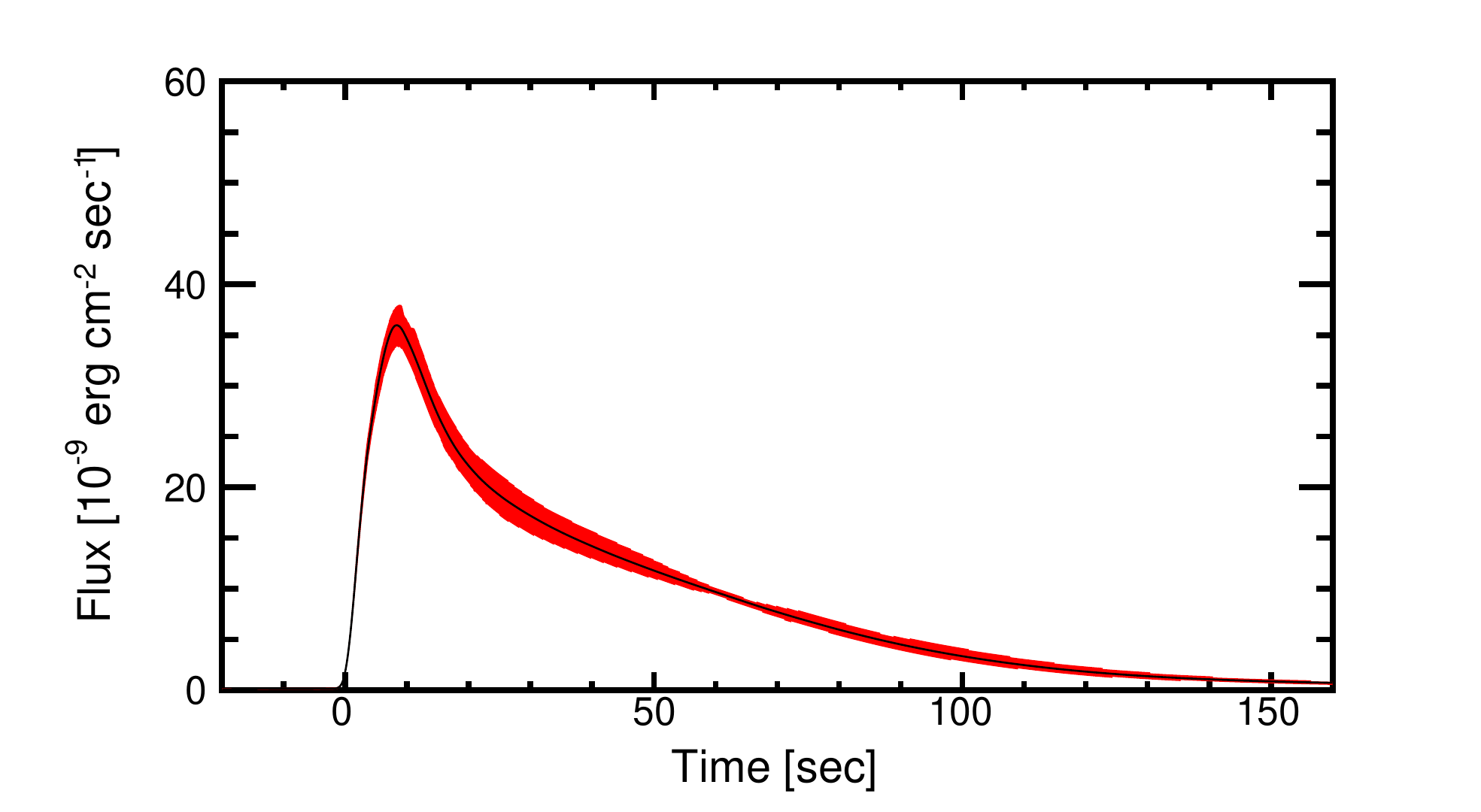}}
 \end{center}
 \caption{Light curve processing steps for the 20 bursts of the
 0.111~$\dot{M}_{\rm{E}}$, $X=0.7$, $Z=0.02$,
 $Q_{\rm{b}}=0.1$~MeV/u, $R=1$ simulation for an arbitrary
 distance and redshift. The first burst (black line in (a)) is not
 included in the averaged light curve and uncertainty band in (c).}
 \label{fig:LCprocess}
\end{figure}

In order to compare to observational data, bursts in a simulated
sequence needed to be stacked (as is done with observed light
curves) so that an average light curve and uncertainty band could be
calculated. The first burst in a sequence was excluded, as the first
simulated burst is typically far more energetic than subsequent
bursts~\citep{Woos04}. The burst start time $t=0$ was defined as the point
when the luminosity crossed a threshold indicating thermonuclear
runaway. Individual bursts were mapped onto the same time grid with
a linear spline and, to mitigate numerical noise, each burst light curve was smoothed by averaging
over the luminosity for a $\pm$1~s time window. This smoothing is
frequently done for multi-zone numerical calculations of
X-ray burst light curves, e.g. as described for the {\tt KEPLER}
models of \citet{Cybu16}\footnote{An example of when this has not
been done are the light curves of \citet{Jose10}, where
(inconsequential) sharp,
unphysical features are seen in some calculation results.}. Using luminosity
data from all bursts (after the first) in a sequence, an average
luminosity and upper and lower $1\sigma$ uncertainties were computed
for each time point. An example of this process is shown in
Fig.~\ref{fig:LCprocess}.

\section{Observational Data}
The observed light curve data for GS 1826-24 are courtesy of the
Multi-Instrument Burst Archive
(MINBAR)\footnote{https://burst.sci.monash.edu/minbar/}. The
data analysis is described in \citet{Gall17} and is briefly
rehashed here. Observational
data from the Rossi X-ray Timing Explorer~\citep{Gall04,Gall08} 
were stacked and averaged in a similar
manner as described above for the simulated light curves of this
work. Since uninterrupted observation was not possible due to
periodic occultation by the Earth, observed recurrence times were
determined using an iterative approach. Each burst was assigned a
trial integer indicating which burst it was in the sequence and a
fit was performed to quantify how well the set of assignments
matched the data if one assumes a regular recurrence time. The
observed $\dot{M}$ were determined using a distance of 6.1~kpc, a bolometric
correction $c_{\rm{bol}}$ between $\approx1.75-1.8$, and the average
persistent flux $F_{\rm{p}}$ over the burst sequence, where
$\dot{M}=4\pi d^{2} F_{\rm{p}}c_{\rm{bol}}/\dot{M}_{\rm{E}}$. These
observed $\dot{M}$ are potentially systematically shifted by some
factor from the true $\dot{M}$ based on the fact that no burst
anisotropy $\xi$~\citep{Fuji88,He16} is assumed. Anisotropy accounts
for the fact that X-ray flux can be beamed toward or away from the
observer, where the effect depends on the accretion disk geometry
and source inclination angle
and can be different for the burst flux and the persistent flux.
This is described in more detail in the discussion.

\section{Model-Observation Comparisons}

Comparison to observations required adjusting the simulated light
curve for the distance and surface gravitational redshift
$(1+z)$. 
The burst anisotropy $\xi_{\rm{b}}$ is included in the distance, 
so that distance is $d\xi_{\rm{b}}^{1/2}$. The
luminosity $L$ to flux $F$ conversion is
$F=L/(4\pi\xi_{\rm{b}}(1+z)d^{2})$~\citep{Gall17}. Time is
redshifted by multiplying the simulation time by $(1+z)$. In
principle, the neutron star mass and radius adopted for the
simulations correspond to $(1+z)=1.26$, so choosing other redshifts
is inconsistent. However, in practice burst properties are
insensitive to modest changes in $M_{\rm{NS}}$ and $R_{\rm{NS}}$
~\citep{Ayas82,Zamf12}.

\begin{figure}[t]
\begin{center}
\includegraphics[width=1.0\columnwidth,angle=0]{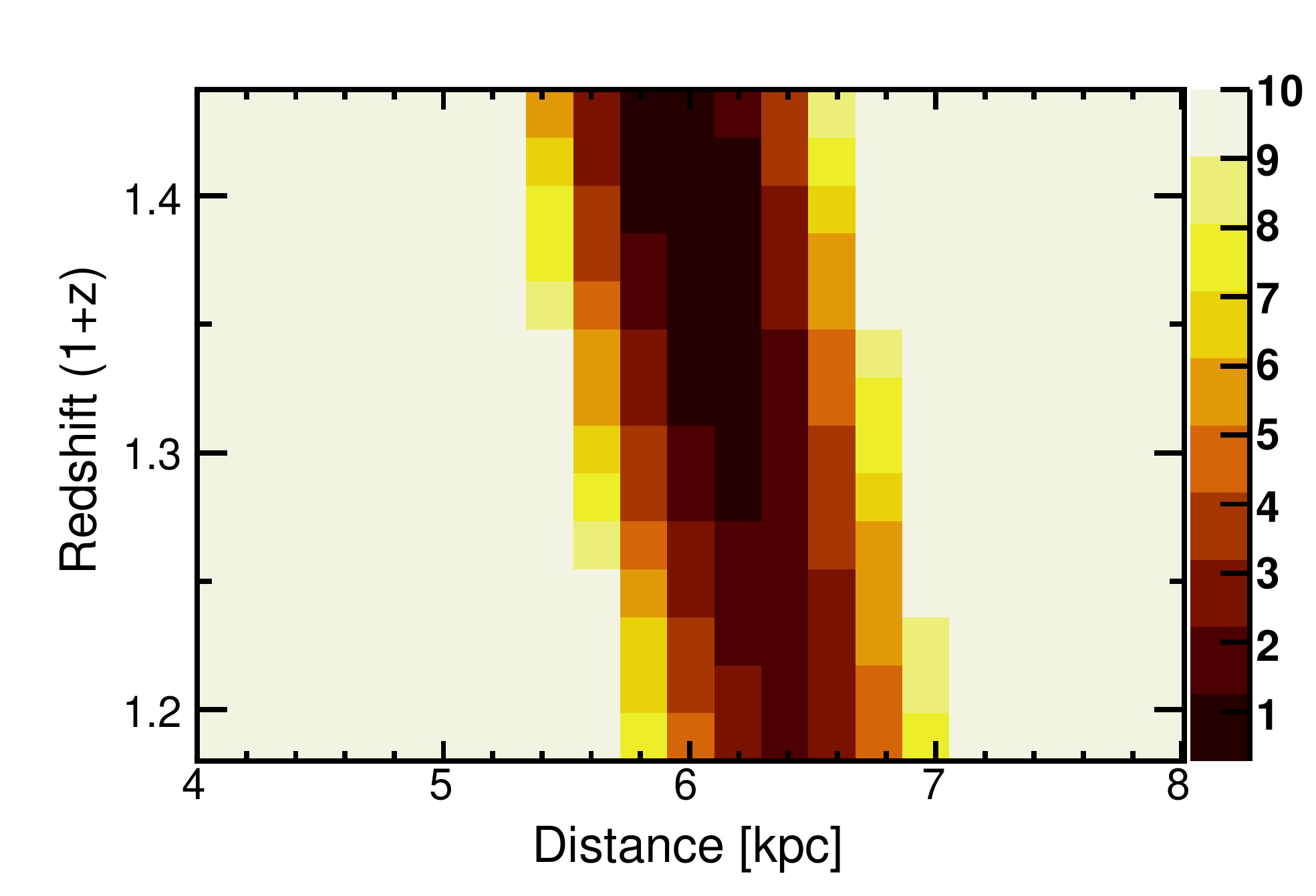}
\caption{$\chi^{2}_{\rm{red}}$, indicated by the
color, for a distance $d\xi_{\rm{b}}^{1/2}$ and redshift, using the optimum
$\delta t$,
for the {\tt MESA} calculation that is the best-fit 
		  to the year 2007 burst epoch for GS 1826-24.
		  $\chi^{2}_{\rm{red}}$$>$$10$ are included in the upper bin.
\label{DZfit}}
\end{center}
\end{figure}

Model-observation comparisons were performed by calculating $F(t)$
for the averaged light curve for the highest $\dot{M}$ in a set of
three and
comparing to the year 2007 burst epoch of GS 1826-24. For
each point in a grid of $d\xi_{\rm{b}}^{1/2}$, $(1+z)$, and time-shift
$\delta t$, $\chi^{2}_{\rm{red}}$ was calculated using data from
 $t=0-50$~s. $\delta t$ is
necessary, so that neither the burst rise nor tail dominates
$\chi^{2}_{\rm{red}}$. $d\xi_{\rm{b}}^{1/2}$ varied in steps of
0.2~kpc from 4 to
8~kpc, based on observational limits~\citep{Gall04}. $(1+z)$ varied
in steps of 0.02 from 1.18 to 1.44, in order to roughly stay within
the range $R_{\rm{NS}}\sim8-15$~km determined by \citet{Stei10} for
$M_{\rm{NS}}=1.4 M_{\odot}$. $\delta t$ varied from 0.5 to 1.5~s in
steps of 0.1~s, as the best-fit was located in this range for each of
the 84 models. The results for
the best-fit out of all models (which also roughly reproduced
the observed $\Delta t_{\rm{rec}}$, see Fig.~\ref{RecVsMdot}), 
is shown in Fig.~\ref{DZfit}. The tight constraints on
$d\xi_{\rm{b}}^{1/2}$
are due to its strong impact on the peak $F$, where as $(1+z)$ is
poorly constrained due to the competition between fitting the burst
rise and burst tail (and is sensitive to the range over which time is
fit)~\citep{Zamf12}.

\begin{figure*}[t]
\begin{center}
\includegraphics[width=0.81\textwidth,angle=0]{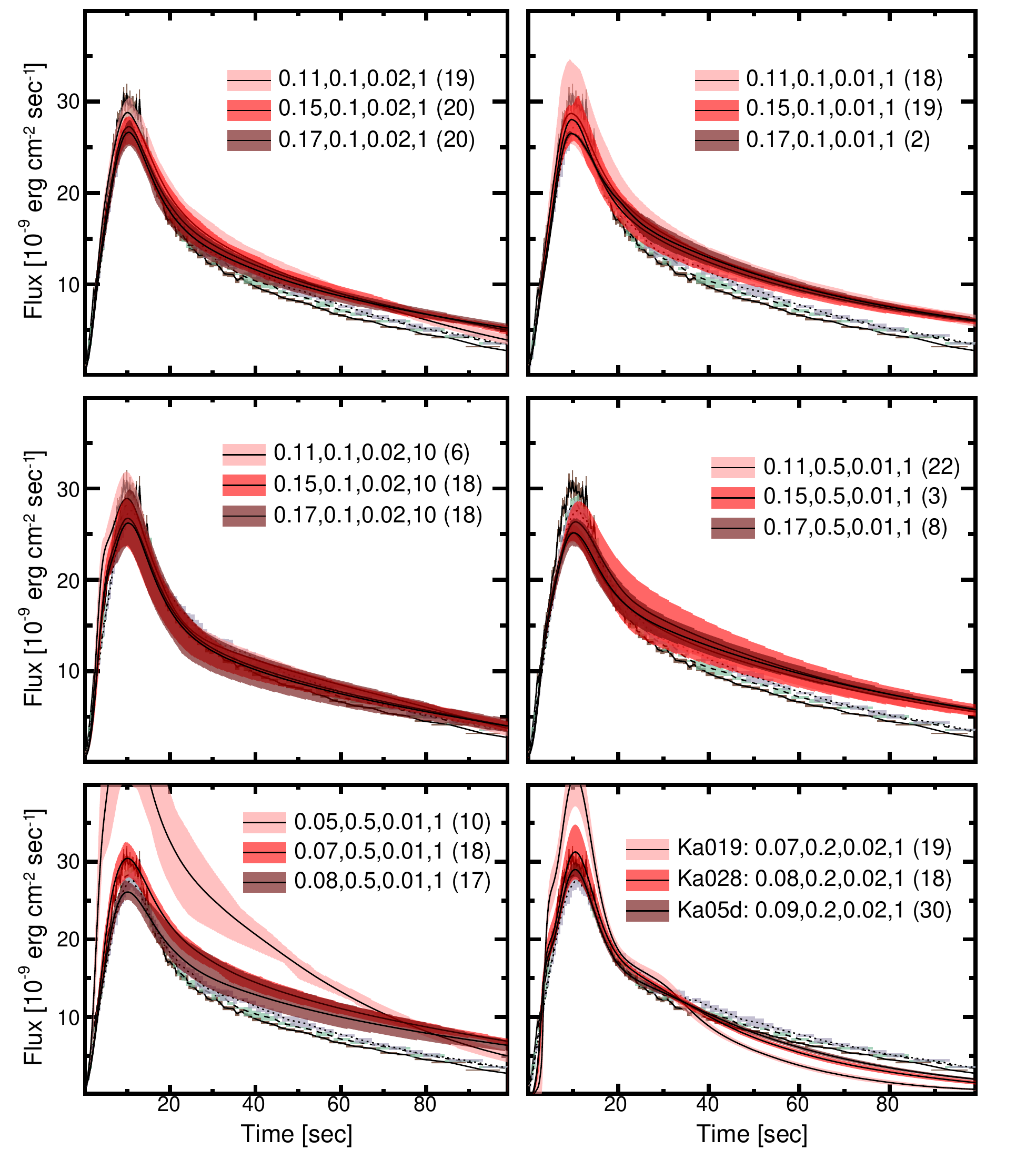}
\caption{Comparison of light curves calculated with
{\tt MESA} (red bands) to GS 1826-24 light curves observed for the
years 1998 (brown), 2000 (green) , and 2007 (purple). Legends
indicate the same information as the legend in
Fig.~\ref{LCexamples}, where the additional number in parentheses is
$N$. The red bands in the lower-right panel are light curves
calculated with {\tt KEPLER} from \citet{Lamp16} using the
optimum distance and redshift determined in \citet{Gall17}.
\label{FitComparisons}}
\end{center}
\end{figure*} 

To move beyond previous studies, a consistent comparison to the year
1998, 2000, and 2007 burst epochs was performed by calculating $F(t)$
for each model using $d\xi_{\rm{b}}^{1/2}$, $(1+z)$, and $\delta t$ obtained
for the best fit to the year 2007 outburst for the highest $\dot{M}$
model in a set of three. I.e. $F(t)$ for $\dot{M}=0.05$ and
$0.07~\dot{M}_{\rm{E}}$ models were calculated using the best-fit
$d\xi_{\rm{b}}^{1/2}$, $(1+z)$, and $\delta t$ found for
$\dot{M}=0.08~\dot{M}_{\rm{E}}$ models, whereas $\dot{M}=0.11$ and
$0.15~\dot{M}_{\rm{E}}$ $F(t)$ were calculated based on
$\dot{M}=0.17~\dot{M}_{\rm{E}}$ models, for the same $Q_{\rm{b}}$,
$Z$, $X$, and $R$. 
The justification for comparing simulations with $\dot{M}$
higher than observed values to observed light curves is that
$\xi_{\rm{b}}$
can differ from the persistent anisotropy $\xi_{\rm{b}}$ in between
bursts, so that the true $\dot{M}$ could be 
different than inferred from observations~\citep{Fuji88,Gall17}. Example
model-observation comparisons are shown in
Fig.~\ref{FitComparisons}, including a comparison to the best-fit
found with {\tt KEPLER}, as reported by \citet{Gall17}, using light curves
calculated by \citet{Lamp16}. The lowest $\dot{M}$ {\tt KEPLER}
model shown is not as low as would be required to match the
observed $\dot{M}$ ratio, but is the closest available.
$\Delta t_{\rm{rec}}$ 
is the average time between thermonuclear runaways using the best-fit
$(1+z)$ for the highest $\dot{M}$ in the set of three. Comparisons
to the observed $\Delta t_{\rm{rec}}$, normalizing $\dot{M}$ so that
the highest $\dot{M}$ of a set of three (i.e. 0.08 or
0.17~$\dot{M}_{\rm{E}}$) matches $\dot{M}$
 for the year 2007 epoch of GS 1826-24, are shown in
Fig.~\ref{RecVsMdot}.

\begin{figure}[t]
\begin{center}
\includegraphics[width=1.0\columnwidth,angle=0]{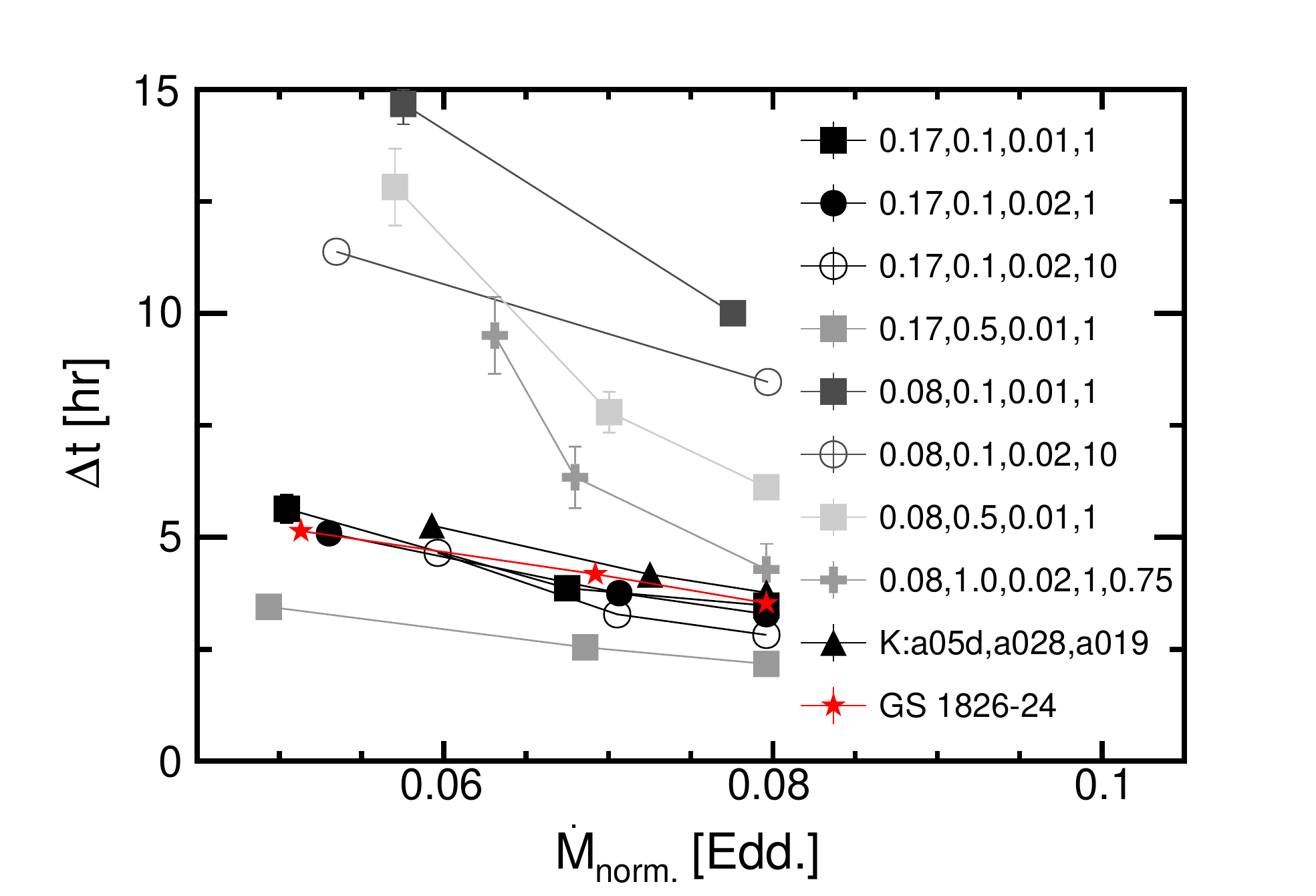}
\caption{Relationship between $\Delta t_{\rm{rec}}$
and normalized accretion rate. Connected symbols indicate bursts
belonging to a triplet of simulated accretion rates.
The legend indicates the same information as in Fig.~\ref{LCexamples}, where
the extra column for the eighth model indicates $X=0.75$.
The {\tt KEPLER} simulations from
Fig.~\ref{FitComparisons} and observed properties for GS 1826-24 are
shown for comparison.
\label{RecVsMdot}}
\end{center}
\end{figure} 

\section{Discussion}

\subsection{Model Parameter Impacts}
Prior to discussing the results from model-observation comparisons, the
impact of model parameters on the X-ray burst light curve and recurrence
time are briefly discussed. 

Increased $\dot{M}$ decreases $\Delta
t_{\rm{rec}}$. This is because the shallow heating scales with the accretion
rate, increased heating speeds up the CNO cycle, which results in an
earlier arrival at the temperature and
He-abundance required to trigger the $3\alpha$ reaction for burst
ignition. Lower $\dot{M}$ therefore requires more H to be burned
prior to burst ignition, resulting in a smaller H/He ratio at burst
ignition, and therefore a relatively He-rich burst. He-rich bursts
burn fuel more rapidly, with higher peak luminosities and
shorter tail decay times as compared to less He-rich
conditions~\citep{Wein06}.

Increased $Q_{\rm{b}}$ decreases $\Delta t_{\rm{rec}}$ for a given
$\dot{M}$ due to the influence of shallow heating on burst
recurrence discussed above for $\dot{M}$. Similarly, for a given
$\dot{M}$, increased $Q_{\rm{b}}$ preserves H prior to burst
ignition, extending the burst tail, which is powered by H-burning.
When considering a
range of $\dot{M}$, increased $Q_{\rm{b}}$ increases the curvature
in the trend for the $\dot{M}$-$\Delta t_{\rm{rec}}$ relationship. This is because
the shallow heating in the model results from the product of
$\dot{M}$ and $Q_{\rm{b}}$, and therefore increasing both 
has a nonlinear influence
on $\Delta t_{\rm{rec}}$.

Increased $Z$ corresponds to increased CNO abundances, increasing
the amount of H-burning prior to burst ignition and leaving less H
to burn during the burst. $Z$ was varied over a relatively small
range here, so the only obvious impact is 
a slightly extended burst tail for
$Z=0.01$ relative to $Z=0.02$. Increased $X$, at the expense of $Y$,
naturally results in
a reduced He abundance at burst ignition, and therefore a decreased peak
luminosity, and more H left to burn during the burst, and therefore
an extended burst tail.

The influence of $R$ on the burst properties derives from the nature
of $^{15}\rm{O}(\alpha,\gamma)$ as a ``valve" controlling the flow
of material out of the hot CNO cycle during interburst
burning~\citep{Fisk06,Cybu16}. Decreasing the
$^{15}\rm{O}(\alpha,\gamma)$ reaction rate (increasing $R$) reduces the
amount of material escaping the hot CNO cycle during quiescent
burning, enabling more He to be produced prior to burst ignition,
shortening $\Delta t_{\rm{rec}}$ and
resulting in a more He-rich burst.

\subsection{$\dot{M}$ and $Q_{\rm{b}}$ Constraints}
Fig.~\ref{FitComparisons} demonstrates that light curve shape for
the year 2007 burst epoch of GS 1826-24 can be accommodated for
several $\dot{M}$ and $Q_{\rm{b}}$, meaning reproduction of a light
curve for a single observed $\dot{M}$ is insufficient to constrain
an X-ray bursting source's conditions with model-observation
comparisons. While $\dot{M}=0.08~\dot{M}_{\rm{E}}$ reproduces the
year 2007 epoch, lower $\dot{M}$ in the set result in a larger peak
flux and shorter burst tail decay than seen in observations,
particularly for the lowest $\dot{M}$. The figure shows that this
result is not only seen with {\tt MESA}, but also for {\tt KEPLER}
models. Therefore, the GS 1826-24 $\dot{M}$ for observed bursting
epochs must be larger than previously inferred.

Fig.~\ref{RecVsMdot} demonstrates that $\Delta t_{\rm{rec}}$
provides an additional necessary discriminant, as models reproducing
the light curve shape for all three observed epochs do not
necessarily reproduce the observed $\Delta t_{\rm{rec}}$. Though
$Q_{\rm{b}}=0.5$~MeV/u can accommodate the light curve shape for all
bursting epochs, $\Delta t_{\rm{rec}}$ is significantly shorter than
for observations. Therefore, shallow heating in GS 1826-24 is limited
to $\leq0.5$~MeV/u, providing an example how multi-epoch X-ray burst
modeling can be used to constrain the shallow heating mechanism in
accreting neutron star outer layers. 

It is evident that  $X=0.75$
cannot be accommodated either, since there is a significant
curvature in $\Delta t_{\rm{rec}}$ for decreasing $\dot{M}$ which is
not seen in the observed data. One sees a similar behavior in {\tt
KEPLER} models, which can be seen by comparing models a003 and a020
of \citet{Lamp16}. $X$ less than $0.7$ were not
explored here as this would move toward the conditions for helium
bursts, as most or all of the hydrogen would be burned stably before
burst ignition.

The best-fit {\tt MESA} model for
light-curve shape and $\Delta t_{\rm{rec}}$ has
$\dot{M}=0.17~\dot{M}_{\rm{E}}$ (for the year 2007 epoch),
$Q_{\rm{b}}=0.1$~MeV/u, $R=1$, $X=0.70$, and $Z=0.02$, though the same
conditions with $Z=0.01$ perform nearly as well.

\subsection{Comparison to {\tt KEPLER}}
Fig.~\ref{RecVsMdot} also highlights
a discrepancy between {\tt MESA} and {\tt KEPLER} models. While
{\tt KEPLER} reproduces the year 2007 epoch $\Delta t_{\rm{rec}}$
with $\dot{M}=0.09~\dot{M}_{\rm{E}}$, {\tt MESA} models require
$\dot{M}=0.17~\dot{M}_{\rm{E}}$, as noted by \citet{Paxt15}.
This cannot be explained by the slightly higher $Q_{\rm{b}}$
employed in the best-fit for {\tt KEPLER}~\citep{Gall17}, as the
{\tt MESA} model with $\dot{M}=0.08~\dot{M}_{\rm{E}}$ and $Q_{\rm{b}}$ 0.5~MeV/u
results in $\Delta t_{\rm{rec}}$ roughly 2/3 larger than
observed for the year 2007 epoch. Systematic comparisons between
{\tt MESA} and {\tt KEPLER}, which are beyond the scope of this
work, are necessary to resolve this
discrepancy. 

Nonetheless, the constraints on $d\xi_{\rm{b}}^{1/2}$
for past {\tt KEPLER} fits and for this work
are in agreement, where the best fit here (see Fig.~\ref{DZfit})
favors 6~kpc and the most recent {\tt KEPLER} results~\citep{Gall17}
favor 6.1~kpc. This work favors a much larger redshift than
\citet{Gall17}, $(1+z)=1.42$ as compared to $1.23$, however
Fig.~\ref{DZfit} demonstrates that $(1+z)$ down to $\sim1.28$
performs nearly as well. As in \citet{Gall17}, uncertainties are not quoted
here due to the large number of systematics which will require
several further studies to quantify. 
It should be noted that $(1+z)=1.42$ corresponds to
$R_{\rm{NS}}=8.2$~km for the canonical $M_{\rm{NS}}=1.4~M_{\odot}$,
which is smaller than expectations~\citep{Stei13}, though
$R_{\rm{NS}}=11.7$~km for $M_{\rm{NS}}=2.0~M_{\odot}$~\citep{Lamp16}.

\subsection{Anisotropies}
The burst anisotropy $\xi_{\rm{b}}$ and persistent anisotropy
$\xi_{\rm{p}}$ between bursts can differ substantially due to the
burst influence on accretion disk geometry. The 
ratio $\xi_{\rm{p}}/\xi_{\rm{b}}$ is determined by the inclination
angle relative to the observer and accretion disk geometry. This ratio can be inferred from
simulation results via
$\xi_{\rm{p}}/\xi_{\rm{b}}=\dot{M}c^{2}(z/(1+z))/(4\pi
d^{2}F_{\rm{p}}c_{\rm{bol}})$, where $c$ is the speed of light~\citep{Hege07}. 
Using this work's best-fit
$\dot{M}$, $z$, and $d$ and $F_{\rm{p}}$ and $c_{\rm{bol}}$ from
\citet{Gall17}, $\xi_{\rm{p}}/\xi_{\rm{b}}=3.5$. This could be
explained (see Fig.~12 of \citet{He16}) by a flat accretion disk 
for a system with a relatively
high inclination angle $\theta\approx80^{\circ}$. For the same
conditions, the {\tt KEPLER} best-fit requires
$\theta\approx65^{\circ}$, whereas roughly the same $\theta$
explains the best-fit from both codes if a curved accretion disk is
assumed. 


\begin{figure}[t]
\begin{center}
\includegraphics[width=1.0\columnwidth,angle=0]{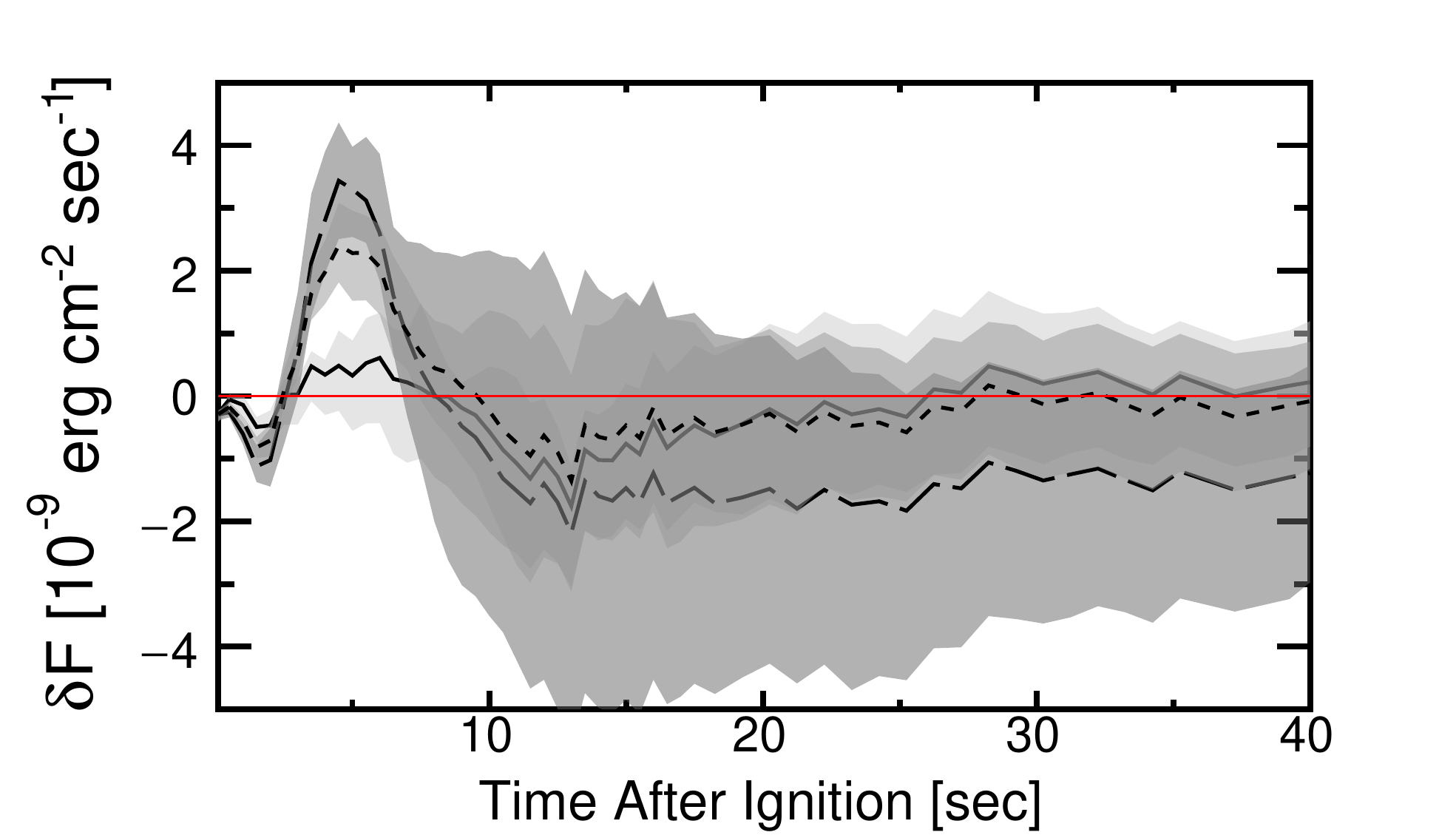}
\caption{$\delta F=F_{\rm{{\tt
MESA}}}-F_{\rm{GS2007}}$
over the early light curve, where the first $\sim$10~s are the light curve
rise, for models with
$\dot{M}=0.17~\dot{M}_{\rm{E}}$, $Q_{\rm{b}}=0.1$~MeV/u, $Z=0.02$,
$X=0.7$,
and, from light to dark bands, $R=1, 5, 10$. 
\label{LCResid}}
\end{center}
\end{figure} 

\subsection{Possibility to Constrain the
$^{15}\rm{O}(\alpha,\gamma)^{19}\rm{Ne}$ Reaction Rate}
It is apparent from Fig.~\ref{RecVsMdot} that $R$ has a relatively
modest impact on $\Delta t_{\rm{rec}}$, in agreement with prior
observations~\citep{Cybu16}. However, as shown in
Figs.~\ref{FitComparisons} and \ref{LCResid}, $R$ significantly increases
the departure from linearity in the light curve rise, known as
the convexity $\mathcal{C}$~\citep{Maur08} (where $\mathcal{C}=0$ is
linear). The year 1998, 2000, and 2007 epochs of GS 1826-24 exhibit a low $\mathcal{C}$, 
whereas {\tt MESA} models with $R>1$ show an increase in
$\mathcal{C}$ due to a shoulder introduced in the light curve rise.
A similar shoulder is present in {\tt KEPLER} models for
$R=10$~\citep{Cybu16}.

It is possible that this signature in the light curve
could be erased by convolving the
one-dimensional results presented here with a more
sophisticated treatment for flame-spreading on the neutron star
surface, which also impacts the light curve rise. 
For instance, \citet{Maur08} found using a phenomenological model
that the longitudinal dependence of the flame speed can result in
$\mathcal{C}>0$ or $<0$ depending on the ignition latitude. Since
$\mathcal{C}\sim0$ for the GS 1826-24 1998, 2000, and 2007 burst
epochs, the $R=5,10$ models presented here could potentially
describe the observational data if convolved with near-polar
burst ignition. Alternatively, relatively slow flame-spreading from
an equatorial ignition could smear-out any intrinsic bump-like
artifacts in the light curve rise; however, \citet{Zamf10} found that
this would require a flame that takes $\sim7$~s to encompass the
neutron star surface, which is several times longer than
inferred from oscillations in the burst light curve rise~\citep{Chak14}.

Taking the current results at face value suggests
that the $^{15}\rm{O}(\alpha,\gamma)$
reaction rate cannot be more than $5\times$ lower than the currently accepted
rate of \citet{Davi11}. This limit is more stringent
than the constraint derived from nuclear physics
experiments, for which the $3\sigma$ uncertainty sets
a lower limit $\gtrsim\times10$~\citep{Davi11}.
This limit for the 
$^{15}\rm{O}(\alpha,\gamma)^{19}\rm{Ne}$
reaction rate, $>5\times$ lower than \citet{Davi11}, 
implies that the $\alpha$-particle decay branching ratio for the key
4.03~MeV resonance in $^{19}\rm{Ne}$ is likely within reach of a newly
developed experimental probe using radioactive ion
beams~\citep{Wred17}.
Nonetheless, it should be
stressed that more reliable constraints will require systematic
investigations, beyond the scope of this work, which employ 
various treatments of effects impacting the
light curve rise that are not included here, especially flame
spreading on the neutron star surface. 
A large number of calculations 
are underway which
will examine the {\tt MESA} X-ray burst model sensitivity to
other nuclear reaction
rates, similar to the study of~\citet{Cybu16}.

\section{Conclusions}

In summary, a large number of X-ray burst model calculations
performed with the
code {\tt MESA} have been used to reproduce the year 1998, 2000, and
2007 bursting epochs from GS 1826-24. It has been shown that
$\dot{M}$ for these bursting epochs must be larger than previously
inferred. This work also shows that model-observation
comparisons for X-ray burst light curves and $\Delta t_{\rm{rec}}$
performed consistently for several $\dot{M}$ are necessary to 
remove model degeneracies. Consistent comparisons can be used to
constrain $Q_{\rm{b}}$ for a bursting source and can possibly
set a lower limit
on the $^{15}\rm{O}(\alpha,\gamma)$ reaction rate. The
$Q_{\rm{b}}<0.5$~MeV/u limit for GS 1826-24 provides a valuable
constraint that can be used to investigate the origins of the poorly
understood shallow heating mechanism in accreting neutron stars. 
Furthermore, using the case of
$^{15}\rm{O}(\alpha,\gamma)$, this work shows that it is possible for
X-ray burst model-observation comparisons to constrain reaction
rates of the $rp$-process, though the constraints determined here
are contingent upon astrophysical effects such as flame spreading
on the neutron star surface.
It is likely that
constraining other nuclear reaction rates will require comparisons to more
observable properties and many more model calculations due to their more subtle 
impacts on the
X-ray burst light curve~\citep{Cybu16}. Logically, such
investigations may be
extended to nuclear masses as well~\citep{Scha17}.

\begin{acknowledgments}
I thank Duncan
Galloway, Zac Johnston, and Hendrik Schatz for useful discussions, and acknowledge
the instructive tutorials 
from Ed Brown, Andrew Cumming, and Rob Farmer located on the
{\tt MESA} Marketplace
({\tt http://cococubed.asu.edu/mesa\_market/}) which helped at the
beginning stages of this work. Relevant {\tt MESA}
inputs for this work are available on the {\tt MESA} Marketplace. 
This work was supported in part by the U.S. Department of Energy
under grant \textnumero~DE-FG02-88ER40387. 
This material is based on
work supported by the National Science Foundation under grant
\textnumero~PHY-1430152 (Joint Institute for Nuclear Astrophysics--Center for the Evolution of the Elements).
\end{acknowledgments}

\bibliographystyle{apj}
\bibliography{ClockReferences}

\begin{thebibliography}{}
\expandafter\ifx\csname natexlab\endcsname\relax\def\natexlab#1{#1}\fi

\bibitem[{Arcones {et~al.}(2017)}]{Arco17}
Arcones, A., {et~al.} 2017, Prog. Part. Nucl. Phys., 94, 1

\bibitem[{Ayasli \& Joss(1982)}]{Ayas82}
Ayasli, S., \& Joss, P.~C. 1982, Astrophys. J., 256, 637

\bibitem[{Bildsten(2000)}]{Bild00}
Bildsten, L. 2000, in Proc. 10$^{th}$ Astrophys. Conf., AIP Conf. Ser., Vol.
  522, 359

\bibitem[{Brown \& Cumming(2009)}]{Brow09}
Brown, E., \& Cumming, A. 2009, Astrophys. J., 698, 1020

\bibitem[{Chakraborty \& Bhattacharyya(2014)}]{Chak14}
Chakraborty, M., \& Bhattacharyya, S. 2014, Astrophys. J., 792, 4

\bibitem[{Cyburt {et~al.}(2016)Cyburt, Amthor, Heger, Johnson, Keek, Meisel,
  Schatz, \& Smith}]{Cybu16}
Cyburt, R.~H., Amthor, A.~M., Heger, A., {et~al.} 2016, Astrophys. J., 830, 55

\bibitem[{Cyburt {et~al.}(2010)}]{Cybu10}
Cyburt, R.~H., {et~al.} 2010, Astrophys. J. Suppl. Ser., 189, 240,
  \url{https://groups.nscl.msu.edu/jina/reaclib/db}

\bibitem[{Davids {et~al.}(2011)Davids, Cyburt, Jos\'{e}, \& Mythili}]{Davi11}
Davids, B., Cyburt, R.~H., Jos\'{e}, J., \& Mythili, S. 2011, Astrophys. J.,
  735, 40

\bibitem[{Fisker {et~al.}(2006)Fisker, G\"{o}rres, Wiescher, \&
  Davids}]{Fisk06}
Fisker, J.~L., G\"{o}rres, J., Wiescher, M., \& Davids, B. 2006, Astrophys. J.,
  650, 332

\bibitem[{Fisker {et~al.}(2008)Fisker, Schatz, \& Thielemann}]{Fisk08}
Fisker, J.~L., Schatz, H., \& Thielemann, F.-K. 2008, Astrophys. J. Suppl.
  Ser., 174, 261

\bibitem[{Fisker {et~al.}(2007)Fisker, Tan, G\"{o}rres, \& Wiescher}]{Fisk07}
Fisker, J.~L., Tan, W., G\"{o}rres, J., \& Wiescher, M. 2007, Astrophys. J.,
  665, 637

\bibitem[{Fujimoto(1988)}]{Fuji88}
Fujimoto, M.~Y. 1988, Astrophys. J., 324, 995

\bibitem[{Galloway {et~al.}(2004)Galloway, Cumming, Kuulkers, Bildsten,
  Chakrabarty, \& Rothschild}]{Gall04}
Galloway, D.~K., Cumming, A., Kuulkers, E., {et~al.} 2004, Astrophys. J., 601,
  466

\bibitem[{Galloway {et~al.}(2017)Galloway, Goodwin, \& Keek}]{Gall17}
Galloway, D.~K., Goodwin, A.~J., \& Keek, L. 2017, Publ. Astron. Soc. Aust.,
  34, 19

\bibitem[{Galloway {et~al.}(2008)Galloway, Muno, Hartman, Dimitrios, \&
  Chakrabarty}]{Gall08}
Galloway, D.~K., Muno, M.~P., Hartman, J.~M., Dimitrios, P., \& Chakrabarty, D.
  2008, Astrophys. J. Suppl. Ser., 179, 360

\bibitem[{Grevesse \& Sauval(1998)}]{Grev98}
Grevesse, N., \& Sauval, A.~J. 1998, Space Sci. Rev., 85, 161

\bibitem[{Gupta {et~al.}(2007)Gupta, Brown, Schatz, M\"{o}ller, \&
  Kratz}]{Gupt07}
Gupta, S., Brown, E.~F., Schatz, H., M\"{o}ller, P., \& Kratz, K.-L. 2007,
  Astrophys. J., 662, 1188

\bibitem[{He \& Keek(2016)}]{He16}
He, C.-C., \& Keek, L. 2016, Astrophys. J., 819, 47

\bibitem[{Heger {et~al.}(2007)Heger, Cumming, Galloway, \& Woosley}]{Hege07}
Heger, A., Cumming, A., Galloway, D.~K., \& Woosley, S.~E. 2007, Astrophys. J.
  Lett., 671, L141

\bibitem[{Henyey {et~al.}(1965)Henyey, Vardya, \& Bodenheimer}]{Heny65}
Henyey, L., Vardya, M.~S., \& Bodenheimer, P. 1965, Astrophys. J., 142, 841

\bibitem[{Johnston {et~al.}(2018)Johnston, Heger, \& Galloway}]{John18}
Johnston, Z., Heger, A., \& Galloway, D.~K. 2018, Monthly Notices of the Royal
  Astronomical Society, sty757

\bibitem[{Jos\'{e} {et~al.}(2010)Jos\'{e}, Moreno, Parikh, \& Iliadis}]{Jose10}
Jos\'{e}, J., Moreno, F., Parikh, A., \& Iliadis, C. 2010, Astrophys. J. Suppl.
  Ser., 189, 204

\bibitem[{Joss(1978)}]{Joss78}
Joss, P.~C. 1978, Astrophys. J. Lett., 225, L123

\bibitem[{Keek \& Heger(2017)}]{Keek17}
Keek, L., \& Heger, A. 2017, Astrophys. J., 842, 113

\bibitem[{Lamb \& Lamb(1978)}]{Lamb78}
Lamb, D.~Q., \& Lamb, F.~K. 1978, Astrophys. J., 220, 291

\bibitem[{Lampe {et~al.}(2016)Lampe, Heger, \& Galloway}]{Lamp16}
Lampe, N., Heger, A., \& Galloway, D.~K. 2016, Astrophys. J., 819, 46

\bibitem[{Maurer \& Watts(2008)}]{Maur08}
Maurer, I., \& Watts, A.~L. 2008, Mon. Not. R. Astron. Soc., 383, 387

\bibitem[{Meisel \& Deibel(2017)}]{Meis17}
Meisel, Z., \& Deibel, A. 2017, Astrophys. J., 837, 13

\bibitem[{Meisel {et~al.}(2016)Meisel, {George}, {Ahn}, {Bazin}, {Brown},
  {Browne}, {Carpino}, {Chung}, {Cole}, {Cyburt}, {Estrad{\'e}}, {Famiano},
  {Gade}, {Langer}, {Mato{\v s}}, {Mittig}, {Montes}, {Morrissey}, {Pereira},
  {Schatz}, {Schatz}, {Scott}, {Shapira}, {Smith}, {Stevens}, {Tan}, {Tarasov},
  {Towers}, {Wimmer}, {Winkelbauer}, {Yurkon}, \& {Zegers}}]{Meis16}
Meisel, Z., {George}, S., {Ahn}, S., {et~al.} 2016, Phys. Rev. C, 93, 035805

\bibitem[{Parikh {et~al.}(2009)Parikh, Jos\'{e}, Iliadis, Moreno, \&
  Rauscher}]{Pari09}
Parikh, A., Jos\'{e}, J., Iliadis, C., Moreno, F., \& Rauscher, T. 2009, Phys.
  Rev. C, 79, 045802

\bibitem[{Parikh {et~al.}(2008)Parikh, Jose, Moreno, \& Iliadis}]{Pari08}
Parikh, A., Jose, J., Moreno, F., \& Iliadis, C. 2008, Astrophys. J. Suppl.
  Ser., 178, 110

\bibitem[{Parikh {et~al.}(2013)Parikh, Jose, Sala, \& Iliadis}]{Pari13}
Parikh, A., Jose, J., Sala, G., \& Iliadis, C. 2013, Prog. Part. Nucl. Phys.,
  69, 225

\bibitem[{Paxton {et~al.}(2011)Paxton, Bildsten, Dotter, Herwig, Lesaffre, \&
  Timmes}]{Paxt11}
Paxton, B., Bildsten, L., Dotter, A., {et~al.} 2011, Astrophys. J. Suppl. Ser.,
  192, 3, \url{www.mesa.sourceforge.net}

\bibitem[{Paxton {et~al.}(2013)Paxton, Cantiello, Arras, Bildsten, Brown,
  Dotter, Mankovich, Montgomery, Stello, \& Timmes}]{Paxt13}
Paxton, B., Cantiello, M., Arras, P., {et~al.} 2013, Astrophys. J. Suppl. Ser.,
  208, 4

\bibitem[{Paxton {et~al.}(2015)Paxton, Marchant, Schwab, Bauer, Bildsten,
  Cantiello, Dessart, Farmer, Hu, \& Langer}]{Paxt15}
Paxton, B., Marchant, P., Schwab, J., {et~al.} 2015, Astrophys. J. Suppl. Ser.,
  220, 15

\bibitem[{Paxton {et~al.}(2018)Paxton, Schwab, Bauer, Bildsten, Blinnikov,
  Duffell, Farmer, Goldberg, Marchant, Sorokina, Thoul, Townsend, \&
  Timmes}]{Paxt18}
Paxton, B., Schwab, J., Bauer, E.~B., {et~al.} 2018, Astrophys. J. Suppl. Ser.,
  234, 34

\bibitem[{Schatz {et~al.}(1999)Schatz, Bildsten, Cumming, \& Wiescher}]{Scha99}
Schatz, H., Bildsten, L., Cumming, A., \& Wiescher, M. 1999, Astrophys. J.,
  524, 1014

\bibitem[{Schatz \& Ong(2017)}]{Scha17}
Schatz, H., \& Ong, W.-J. 2017, Astrophys. J., 844, 139

\bibitem[{{Schatz} {et~al.}(1998)}]{Scha98}
{Schatz}, H., {et~al.} 1998, Phys. Rep., 294

\bibitem[{Schatz {et~al.}(2001)Schatz, Aprahamian, Barnard, Bildsten, Cumming,
  Oullette, Rauscher, Thielemann, \& Wiescher}]{Scha01}
Schatz, H., Aprahamian, A., Barnard, V., {et~al.} 2001, Phys. Rev. Lett., 86,
  3471

\bibitem[{Steiner {et~al.}(2010)Steiner, Lattimer, \& Brown}]{Stei10}
Steiner, A.~W., Lattimer, J.~M., \& Brown, E.~F. 2010, Astrophys. J., 722, 33

\bibitem[{Steiner {et~al.}(2013)Steiner, Lattimer, \& Brown}]{Stei13}
---. 2013, Astrophys. J. Lett., 765, L5

\bibitem[{Tan {et~al.}(2007)Tan, Fisker, G\"{o}rres, Couder, \&
  Wiescher}]{Tan07}
Tan, W.~P., Fisker, J.~L., G\"{o}rres, J., Couder, M., \& Wiescher, M. 2007,
  Phys. Rev. Lett., 98, 242503

\bibitem[{Turlione {et~al.}(2015)Turlione, Aguilera, \& Pons}]{Turl15}
Turlione, A., Aguilera, D.~N., \& Pons, J.~A. 2015, Astron. \& Astrophys., 577,
  A5

\bibitem[{Wallace \& Woosley(1981)}]{Wall81}
Wallace, R.~K., \& Woosley, S.~E. 1981, Astrophys. J. Suppl. Ser., 45, 389

\bibitem[{Weaver {et~al.}(1978)Weaver, Zimmerman, \& Woosley}]{Weav78}
Weaver, T.~A., Zimmerman, G.~B., \& Woosley, S.~E. 1978, Astrophys. J., 225,
  1021

\bibitem[{Weinberg {et~al.}(2006)Weinberg, Bildsten, \& Schatz}]{Wein06}
Weinberg, N., Bildsten, L., \& Schatz, H. 2006, Astrophys. J., 639, 1018

\bibitem[{Wiescher {et~al.}(1999)Wiescher, G\"{o}rres, \& Schatz}]{Wies99}
Wiescher, M., G\"{o}rres, J., \& Schatz, H. 1999, J. Phys. G., 25, R133

\bibitem[{{Woosley} {et~al.}(2004){Woosley}, {Heger}, {Cumming}, {Hoffman},
  {Pruet}, {Rauscher}, {Fisker}, {Schatz}, {Brown}, \& {Wiescher}}]{Woos04}
{Woosley}, S.~E., {Heger}, A., {Cumming}, A., {et~al.} 2004, Astrophys. J.
  Suppl. Ser., 151, 75

\bibitem[{Wrede {et~al.}(2017)}]{Wred17}
Wrede, C., {et~al.} 2017, Phys. Rev. C, 96, 032801(R)

\bibitem[{Zamfir(2010)}]{Zamf10}
Zamfir, M. 2010, Master's thesis, McGill University

\bibitem[{Zamfir {et~al.}(2012)Zamfir, Cumming, \& Galloway}]{Zamf12}
Zamfir, M., Cumming, A., \& Galloway, D.~K. 2012, Astrophys. J., 749, 69

\end{thebibliography}

\end{document}